\documentclass[prd,aps,twocolumn,nofootinbib]{revtex4}%nofootinbib,%onecolumn,%groupedaddress]{revtex4}
\usepackage{epsfig}
\usepackage{amsfonts}
\usepackage{rotating}
\usepackage{sparticles}
\usepackage{dcolumn}
\usepackage{bm}
\usepackage{hyperref}
\hypersetup{
    bookmarks=true,         % show bookmarks bar?
    unicode=false,          % non-Latin characters in Acrobat’s bookmarks
    pdftoolbar=true,        % show Acrobat’s toolbar?
    pdfmenubar=true,        % show Acrobat’s menu?
    pdffitwindow=false,     % window fit to page when opened
    pdfstartview={FitH},    % fits the width of the page to the window
    pdftitle={My title},    % title
    pdfauthor={Author},     % author
    pdfsubject={Subject},   % subject of the document
    pdfcreator={Creator},   % creator of the document
    pdfproducer={Producer}, % producer of the document
    pdfkeywords={keyword1} {key2} {key3}, % list of keywords
    pdfnewwindow=true,      % links in new window
    colorlinks=true,       % false: boxed links; true: colored links
    linkcolor=red,          % color of internal links
    citecolor=cyan,        % color of links to bibliography
    filecolor=magenta,      % color of file links
    urlcolor=green,           % color of external links
    linktocpage=true
}
\usepackage{amsmath,amssymb,amsthm,graphicx,latexsym}
\usepackage{subfigure}
\usepackage{pstricks}
\usepackage{color}

\usepackage{mathrsfs}

\newcommand{\be}{\begin{equation}}
\newcommand{\ee}{\end{equation}}
\newcommand{\bea}{\begin{eqnarray}}
\newcommand{\eea}{\end{eqnarray}}
\newcommand{\bml}{\begin{subequations}}
\newcommand{\eml}{\end{subequations}}
\newcommand{\bfig}{\begin{figure}}
\newcommand{\efig}{\end{figure}}

\begin{document}

\title{ Low \& High scale MSSM inflation, gravitational waves and constraints from Planck }

\author{Sayantan Choudhury$^1$, Anupam Mazumdar$^2$ and Supratik Pal$^{1}$}

\affiliation{$^{1}$Physics and Applied Mathematics Unit, Indian Statistical Institute, 203 B.T. Road, Kolkata 700 108, India\\
$^2$Consortium for Fundamental Physics, Physics Department, Lancaster University, LA1 4YB, UK}

%\vspace{5ex}
\date{\today}
\begin{abstract}
In this paper we will analyze generic predictions of an {\it inflection-point} model of inflation with 
Hubble-induced corrections and study them in light of the {\it Planck} data. Typically inflection-point models 
of inflation can be embedded within Minimal Supersymmetric Standard Model (MSSM)
where inflation can occur below the Planck scale. The flexibility of the potential allows us to match the observed 
amplitude of the TT-power spectrum of the cosmic microwave background radiation with low and high multipoles, spectral tilt, 
and virtually  mild running of the spectral tilt, which can put a bound on an upper limit on the tensor-to-scalar ratio, $r \leq 0.12$.
Since the inflaton within MSSM carries the Standard Model charges, therefore it  is the minimal model of inflation 
beyond the Standard Model which can reheat the universe with the right thermal degrees 
of freedom without any {\it dark-radiation}.

\end{abstract}

\maketitle

\section{\bf Introduction}
\label{intro}

The observational success of primordial inflation arising from the cosmic microwave background (CMB) 
radiation~\cite{Planck-infl,WMAP} has lead to an outstanding question how to embed the inflationary paradigm
within a particle theory~\cite{infl-rev}. Since inflation dilutes all matter except for the quantum vacuum fluctuations 
of the inflaton, it is pertinent that end of inflation creates all the relevant Standard Model degrees of freedom for 
the success of Big Bang Nucleosynthesis~\cite{BBN}, without any extra relativistic degrees of freedom, 
i.e. dark radiation~\cite{Planck-1}~\footnote{Embedding the last 50 -60 e-foldings of inflation within string theory has a major disadvantage.
Due to large number of hidden sectors arising from any string compactifications, it is likely that the inflaton energy density will get dumped 
into the hidden sectors instead of the visible sector~\cite{Cicoli:2010ha}. The branching ratio for the inflaton decay into the visible sector is 
very tiny, therefore reheating the Standard Model degrees of freedom is one of the biggest challenges for any string motivated models of inflation.
Furthermore, many of the compactifications generically lead to extra dark radiation (massless axions) which are already at the verge of being 
ruled out by the present data~\cite{Angus:2013zfa}.}.

This immediately suggests that the inflationary vacuum cannot be arbitrary and the inflaton must decay {\it solely} into the 
Standard Model degrees of freedom. Furthermore, the recent Planck data~\cite{Planck-1} indicates that the perturbations in the
 baryons and the cold dark matter are adiabatic in nature, it is evident that there must be a single source of 
perturbations which is responsible for seeding the fluctuations in all forms of matter~\footnote{In principle more than one fields can still
participate during inflation, but they must do so in such a way that here exists an attractor solution which would yield solely adiabatic perturbations 
and no isocurvature perturbations, such as in the case of assisted inflation~\cite{Assist}.}. 

Strictly speaking this can happen {\it only} if the 
inflaton itself carries the Standard Model charges as in the case of Minimal Supersymmetric Standard Model (MSSM) flat-directions~\cite{Enqvist:2003gh}, where the lightest supersymmetric particle could be the dark matter candidate and can be created from thermal annihilation of the MSSM degrees of freedom. MSSM inflation was first discussed in Refs.~\cite{Allahverdi:2006iq,Allahverdi:2006we,Allahverdi:2006cx}, and in recently, see Refs.~\cite{Wang:2013hva,Mazumdar:2011ih,Choudhury:2011jt,Chatterjee:2011qr,Aulakh:2012st,Dutta}.  The inflaton candidates are made up of
{\it gauge invariant } combinations of squarks (supersymmetric partners of quarks) and sleptons (supersymmetric partners of leptons).

One of the {\it key} ingredients for  embedding inflation within MSSM is that the inflaton VEV must be below the Planck scale, 
$M_{PL}=2.4\times 10^{18}$~GeV. This justifies the application of an effective field theory treatment at low energies.
It is well-known that the potential for the MSSM flat-direction inflaton has high degree of  {\it flexibility} -- the potential can accommodate
{\it inflection-point} below the Planck VEV, which allows a rich class of flat potentials, with a vanishing effective mass,
which has been studied analytically and numerically~\cite{Allahverdi:2006we,Bueno Sanchez:2006xk,Enqvist:2010vd}. The application of 
inflection point inflation is not just limited to particle theory, but such potentials have also found their applications in string theory~\cite{Jain:2008dw}.
 
 It has also been known that  inflection-point models of inflation can occur for a wide range of Hubble values, $H_{inf}$, 
 ranging from $10^{-1}~{\rm GeV}< H_{inf}\leq 10^{13}$~GeV. For very high scale inflation there is a possibility of obtaining 
 signatures for the primordial gravitational waves, namely the B-modes~\cite{Hotchkiss:2011gz} below the Planck scale~\footnote{We are assuming that 
 the gravitational modes can be quantized with quantum initial conditions. If the gravity waves behave classically, then the amplitude
 of the gravitational waves in a simple scalar field driven model will be absolutely zero~\cite{Ashoorioon:2012kh}, as there is no 
 source term for the gravitational waves. }.

The aim of this paper is to show explicitly how large scale inflection-point inflation can match the current CMB 
observables, namely the TT-part of the temperature anisotropy spectrum, low and high multipoles, spectral tilt, 
running of the spectral tilt, and running of running of the spectral tilt. We will also provide the ranges of tensor-to-scalar ratio $r$,
which is compatible with all the data sets~\footnote{Similar studies were undertaken for hybrid inflation model~\cite{Linde:1993cn,Dvali:1994ms},
after the {\it Planck} data release, see~\cite{Pallis:2013dxa}. For power law inflation, see~\cite{Unnikrishnan:2013vga}.}.

In section~\ref{sec-2}, we will discuss the generality of inflection-point potential. In  section\ref{H-sect}, we will provide a brief discussion on supergravity corrections, and the two regimes of inflation will be discussed in section~\ref{2-r}.
In section~\ref{sec-3}, we will discuss 
the cosmological observables relevant for CMB, in section \ref{TRH-sec}, we will discuss reheat temperature and the number of e-foldings. 
In section \ref{tensor}, we will discuss tensor-to-scalar ratio of high and low scale models of inflation. In section~\ref{sec-4}, we will provide the TT-power spectrum (low and high multipoles) for a particular realisation,
and in section~\ref{sec-5}, we will conclude our results.

%%%%%%%%%%%%%%%%%%%%%%%%%%%%%%%%%%%%%%%%%%%%%%%%%%%%%%%%%%%%%%%%%%%%%%%%%%%%%%%%%%%%%%%%%%%%%%%%%%%%%%%%%%%%%%%%%%%%%%%%%%%%%%%%%%%%%%%%%%%%%%%%%%%%%%%%%%%%%

\section{\bf Flat potential around the Inflection-Point }
\label{sec-2}

The most generic  inflection-point potential can be recast as~\cite{Enqvist:2010vd,Hotchkiss:2011gz}:
\begin{equation}\label{rt1}
V(\phi)=\alpha+\beta(\phi-\phi_{0})+\gamma(\phi-\phi_{0})^{3}+\kappa(\phi-\phi_{0})^{4}+\cdots\,,
   \end{equation}
where any generic potential, $V(\phi)$, has been expanded around the inflection-point, $\phi_0$, where $\alpha$ denotes the cosmological 
constant, and coefficients $\beta,~\gamma,~\kappa$ determine the shape of the potential in terms of the model parameters. Typically,
$\alpha$ can be set to zero by fine tuning, but here we wish to keep this term for generality. Note that not all of the coefficients are 
independent once we prescribe inflaton within MSSM.

 %%%%%%%%%%%%FUGURES%%%%%%%%%%%%%%%%%%%%%  
   
\begin{figure}[ht]
\centering
\subfigure[~Shape of the potential. Note the existence of a flat plateau below the Planck scale. ]{
   \includegraphics[width=8cm,height=6cm] {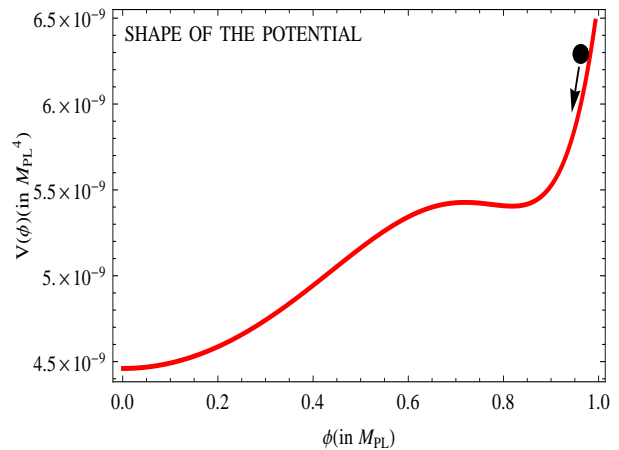}
    \label{fi1}
}
\subfigure[~The blow up picture of the potential near the inflection point. Here $x=\phi-\phi_0$, where $\phi_0$ is the inflection-point.]{
    \includegraphics[width=8cm,height=6cm] {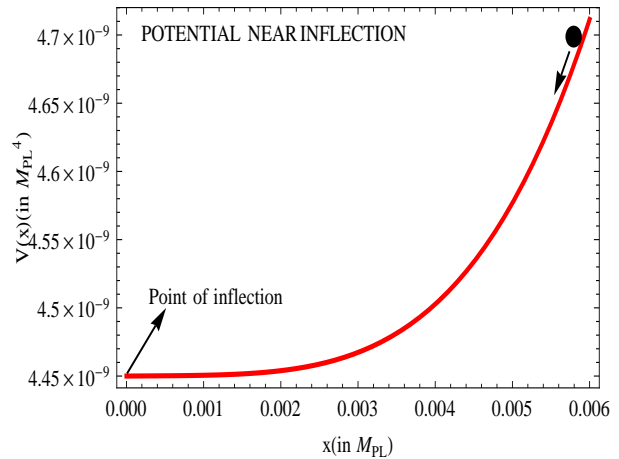}
    \label{fi2}
}
\caption{Inflationary potential including Hubble induced corrections.}\label{fig1}
\end{figure}

%%%%%%%%%%%%%%%%%%%%%%%%%%%%%%%%%%%%%%%%%%%%%%%%%%%%%%%

In Refs.~\cite{Allahverdi:2006iq,Allahverdi:2006we,Allahverdi:2007vy,Boehm:2012rh,Wang:2013hva} the authors have recognized two
 $D$-flat directions which can be the ideal inflaton candidates. Both $\widetilde{u}\widetilde{d}\widetilde{d}$, where
 $\widetilde u,~\widetilde d$ correspond to the right handed squarks, and $\widetilde{L}\widetilde{L}\widetilde{e}$, 
where $\widetilde L$ is the left handed slepton, and $\widetilde e$ is the right handed (charged) leptons,
 flat directions are lifted by higher order superpotential terms of the following simple form: 
\begin{equation} \label{supot}
W (\Phi)= {\lambda \over 6}{\Phi^6 \over M^3_{PL}}\, ,
\end{equation}
 where $\lambda \sim {\cal O}(1)$ coefficient. The scalar component of $\Phi$ superfield, denoted by $\phi$, is given by~
\begin{equation} \label{infl}
\phi = {\widetilde{u} + \widetilde{d} + \widetilde{d} \over \sqrt{3}} ~ ~ ~ , ~ ~ ~ \phi = {\widetilde{L} + \widetilde{L} + 
\widetilde{e} \over \sqrt{3}},
\end{equation}
for the $\widetilde{u}\widetilde{d}\widetilde{d}$ and $\widetilde{L}\widetilde{L}\widetilde{e}$ flat directions, respectively.

%%%%%%%%%%%%%%%%%%%%%%%%%%%%%%%%%%%%%%%%%%

\subsection{Brief discussion on Hubble induced corrections}\label{H-sect}

In addition to Eq.~(\ref{supot}), there are many possible contributions to the vacuum energy. It is conceivable that at high energies the universe is dominated by a large cosmological constant arising from a string theory landscape~\cite{Douglas:2006es}. Our own patch of the universe could be locked in a false vacuum within an MSSM landscape~\cite{Allahverdi:2008bt}, or there could be hidden sector contributions~\cite{Enqvist:2007tf,Lalak:2007rsa}, or there could be a combination of these effects. For simplicity we may attribute such a vacuum energy to a hidden sector. A hybrid model of inflation~\cite{Linde:1993cn} also provides a source of vacuum energy density during inflation. For the purpose of illustration, we will model earlier phases of inflation driven by the superpotential of type:
\begin{equation}
W(S)=M^{2}S\,,
\end{equation}
where $M$ is some high scale which dictates the initial vacuum energy density, and $S$ is the hidden sector superfield. 
The total K\"ahler potential can be of the form \cite{Dine:1995kz,Dine:1995uk,Kasuya:2006wf}:
\begin{equation}
 K=S^{\dagger}S+\Phi^{\dagger}\Phi +\delta K, 
 \end{equation}
where the non-minimal term $\delta K$ can be any of these functional forms:
\begin{equation}
\delta K=f(\Phi^{\dagger}\Phi,S^{\dagger}S)\,,~f(S^{\dagger}\Phi\Phi)\,,~f(S^\dagger S^\dagger\Phi\Phi)\,,~f(S\Phi^{\dagger}\Phi)\nonumber
 \end{equation}
 We will always treat the fields $s,~\phi\ll M_{PL}$. The higher order corrections to the K\"ahler potential are extremely hard to compute. It has been done within a string theory setup~\cite{Berg} but only in very special circumstances, and not for MSSM fields. 
The scalar potential in ${\cal N}=1$ supergravity can be written in terms of superpotential, 
$W$, and K\"ahler potential, $K$, as
\begin{eqnarray}
&  & V=e^{K(\Phi, \Phi^{\dagger})/M_{PL}^2}\Big[ \big( D_{\Phi_i} W(\Phi) \big)
K^{\Phi_i \bar{\Phi}_j} \big(D_{\bar{\Phi}_j} W^*(\Phi^{\dagger}) \big)  \nonumber \\
& & \qquad\qquad\qquad - \frac{3}{M_{PL}^2} \left| W(\Phi) \right|^2 \Big] + ({\rm D-terms}),
\end{eqnarray}
where $F_\Phi \equiv D_{\Phi} W = W_{\Phi} + K_{\Phi} W/M_{PL}^2$, and 
$K^{\Phi_i \bar{\Phi}_j}$ is the inverse matrix of $K_{\Phi_i \bar{\Phi}_j}$, and the subscript denotes derivative with respect to the field. Hereafter, 
we neglect the contribution from the D-term, since the MSSM inflatons are D-flat directions. In the above potential $W=W(S)+W(\Phi)$.

After minimizing the potential along the angular direction, $\theta$ ($\Phi$ = $\phi e^{i \theta}$), we take  the real part 
of $\phi$ by rotating it to the corresponding angle $\theta$, resulting in~\cite{Dine:1995kz,Dine:1995uk,Kasuya:2006wf}:
\begin{eqnarray}\label{h1}
 V(\phi,\theta) &=&V_{0}+\frac{(m^2_\phi+c_{H}H^{2})}{2}|\phi|^{2}\nonumber \\
&& +(a_{H} H +a_\lambda  m_{\phi})\frac{\lambda\phi^{6}} {6M^{3}_{PL}}\cos(6\theta+\theta_{a_{H}}+\theta_{a_\lambda})\nonumber \\
&& ~~~~~~~~~~~~~~~~~~~~~~~~~~~~~~~~~+  \frac{\lambda^{2}|\phi|^{10}}{M^{6}_{PL}}
   \end{eqnarray}
where  the cosmological constant will be determined by the overall inflationary potential,  $V_0\approx 3H^2M_{PL}^2$.
Usually this bare cosmological term can be set to zero from the beginning by tuning the graviton mass. We will consider 
scenarios, where we will have $V_0\neq 0$ and $V_0=0$.

Note that $m_\phi$ and 
$a_\lambda$ are soft-breaking mass and the non-renormalizable $A$-term respectively ($A$ is a positive quantity 
since its phase is absorbed by a redefinition of $\theta$ during the process)~\footnote{The masses are given by:
$$m^2_{\phi}=\frac{m^2_{\widetilde L}+m^2_{\widetilde L}+m^2_{\widetilde e}}{3},~~~
m^2_{\phi}=\frac{m^2_{\widetilde u}+m^2_{\widetilde d}+m^2_{\widetilde d}}{3}$$ for 
$\widetilde L\widetilde L\widetilde e$ and $\widetilde u\widetilde d\widetilde d$ directions respectively. Typically these
masses are set by the scale of supersymmetry, in the low scale case the masses will be typically of order ${\cal O}(1)$~TeV.}.
 The potential also obtains Hubble-induced corrections, with coefficients $c_{H},~a_{H}\sim {\cal O}(1)$. Their exact numerical values
 will depend on the nature of K\"ahler corrections and compactification, which are hard to compute for a generic scenario~\cite{Berg}, 
 but the corrections typically yield $\sim {\cal O}(1)$ coefficients.  The non-renormalizable terms have a  
 periodicity of $2\pi$ in $(\phi,~\theta)$ 2D plane, $\theta_{a_{H}},~\theta_{a_\lambda}$ are the extra phase factors.
 
 %%%%%%%%%%%%%%%%%%%%%%%%%%%%%%%%%%%%%%%%%%%%%%%%%%%%%%%
 
 \section{Two regimes of inflation: Low \& High}\label{2-r}
 
There are two regimes where one can describe  the dynamics of the potential:
 
\subsection{$m_\phi \gg H$: Low scale inflation}\label{01}
 
Since $c_H, a_H~\sim {\cal O}(1)$ the Hubble-induced terms do not play any crucial role in this case, and the scale of inflation remains very low. 
As a result the tensor-to scalar ratio, $r$, becomes too small to be ever detectable. This was the scenario studied in Refs.~\cite{Allahverdi:2006we,Allahverdi:2006iq}. 
% In the low scale scenario, the value of $V_0\leq m_{\phi}^2\phi_0^2$, is negligible and does not contribute to the dynamics.
In this case we can set its value to $V_0=0$ from the beginning by tuning the gravitino mass~\cite{Nilles:1983ge}.
The potential can be minimized along the $\theta$ direction, which reduces to~\cite{Allahverdi:2006iq,Allahverdi:2006we}:
\begin{equation}
V(\phi,\theta) =\frac{m^2_\phi}{2}|\phi|^{2}-a_\lambda  m_{\phi}\frac{\lambda\phi^{6}} {6M^{3}_{PL}}
+  \frac{\lambda^{2}|\phi|^{10}}{M^{6}_{PL}}
\end{equation}
For,
 \begin{equation} \label{dev}
{a_\lambda^2 \over 40} \equiv 1 - 4 \delta^2\, ,
\end{equation}
and $\delta^2 \ll 1$, there exists a point of inflection ($\phi_0$) in $V(\phi)$, where
\begin{eqnarray}\label{vev}
&&\phi_0 = \left({m_\phi M^{3}_{PL}\over \lambda \sqrt{10}}\right)^{1/4} + {\cal O}(\delta^2) \, , \label{infvev} \\
&&\, \nonumber \\
&&V^{\prime \prime}(\phi_0) = 0 \, , \label{2nd}
\end{eqnarray}
at which
\begin{eqnarray}
\label{pot}
\alpha &=&V(\phi_0) = \frac{4}{15}m_{\phi}^2\phi_0^2 + {\cal O}(\delta^2) \, , \\
\label{1st}
\beta&=&V'(\phi_0) = 4 \alpha^2 m^2_{\phi} \phi_0 \, + {\cal O}(\delta^4) \, , \\
\label{3rd}
\gamma&=&V^{\prime \prime \prime}(\phi_0) = 32\frac{m_{\phi}^2}{\phi_0} + {\cal O}(\delta^2) \, .
\end{eqnarray}
The potential is specified completely by $m_\phi$ and $\lambda$. However $m_\phi$ is determined by the soft-SUSY breaking
mass parameter,  which is well constrained by the current ATLAS~\cite{ATLAS} and CMS~\cite{CMS} data,  and we shall take $m_\phi=1$~TeV. 
For $m_\phi\sim 1$~TeV, $H\ast\sim 0.1$~GeV, and our assumption of neglecting $H$ in such a case is well justified.
We will always consider $\lambda =1$ in our analysis.

%%%%%%%%%%%%%%%%%%%%%%%%%%%%%%%%%%%%%%%%%%%%%%%%%%%%%

 \subsection{ $H \gg m_\phi$: High scale inflation}\label{02}
 
  The supergravity corrections become important, the Hubble-induced terms dominate the potential. This can happen
 quite naturally if there exists a previous source of effective cosmological constant term described in Refs.~\cite{Mazumdar:2011ih,Hotchkiss:2011gz}. 
 In this case one can safely ignore the soft SUSY breaking mass term, and since $a_\lambda\sim {\cal O}(1)$, one can 
 safely consider only the Hubble-induced non-renormalizable term. One advantage of considering such a potential is to 
 obtain large tensor-to-scalar ratio, $r$,  which can be within the range of Planck and other future CMB B-mode polarization experiments. We will
 keep $V_0$ in this case, and  the potential simplifies to~\cite{Mazumdar:2011ih}:
\begin{equation}\label{h1a}
 V(\phi)=V_{0}+\frac{c_{H}H^{2}}{2}|\phi|^{2}-\frac{a_{H}H \phi^{6}}{6M^{3}_{PL}}+\frac{|\phi|^{10}}{M^{6}_{PL}}.
\end{equation}
where we have taken $\lambda=1$.
The potential admits inflection point for  $a_H^2\approx40 c_H^2$. We characterize the required fine-tuning by the 
quantity, $\delta$, defined as~\cite{Allahverdi:2006we}
\begin{equation}
\label{newbeta}
\frac{a_H^2}{40c_H^2} = 1-4\delta^2\,.
\end{equation}
When $\vert\delta\vert$ is small, a point of inflection $\phi_0$ exists such that $V^{\prime\prime}\left(\phi_0\right) =0$, with
\begin{equation}
\label{phi0}
\phi_0 = \left(\sqrt{\frac{c_H}{10}} H M_{PL}^{3}\right)^{{1}/{4}}\,.
\end{equation}
For $\delta <1$, we can Taylor-expand the inflaton potential around the inflection point $\phi=\phi_{0}$ similar to 
 Eq.~(\ref{rt1}), where the coefficients are now given by:
\begin{eqnarray}\label{p1}
     \alpha&=&V(\phi_{0})=V_{0}+\left(\frac{4}{15}+\frac{4}{3}\delta^{2}\right)c_{H}H^{2}\phi^{2}_{0}+{\cal O}(\delta^{4}),\\
 \beta&=&V^{'}(\phi_{0})=4\delta^{2}c_{H}H^{2}\phi_{0}+{\cal O}(\delta^{4}),\\
 \gamma&=&\frac{V^{'''}(\phi_{0})}{3!}=\frac{c_{H}H^{2}}{\phi_{0}}\left(32-80\delta^{2}\right)+{\cal O}(\delta^{4}),\\
 \kappa&=&\frac{V^{''''}(\phi_{0})}{4!}=\frac{c_{H}H^{2}}{\phi^{2}_{0}}\left(384-1260\delta^{2}\right)+{\cal O}(\delta^{4}).
   \end{eqnarray}
Note that once we specify $c_H$ and $H$, all the terms in the potential are determined. In this regard the potential 
indeed simplifies a lot to study the cosmological observables. In section~\ref{sec-4}, we will be scanning over $c_H,~H$ 
to obtain the best fits with the current observations.

One must also ensure that the vacuum energy density which generated the
 large cosmological constant in the first place vanishes by the end of slow-roll inflation.
 This typically happens in the case of hybrid inflation~\cite{Linde:1993cn,Dvali:1994ms}, and
 as discussed in~\cite{Enqvist:2010vd,Hotchkiss:2011gz,Burgess:2005sb}. In the string
 landscape~\cite{Douglas:2006es}, or in the case of MSSM~\cite{Allahverdi:2008bt}, this
 can happen through bubble nucleation, provided the rate of nucleation is such
 that $\Gamma_{nucl}\gg H$. In the latter case all the bubbles will belong to the
 MSSM vacuum---similar to the first order phase transition in the electroweak symmetry
 breaking scenario. However, one has to make sure that the cosmological
 constant disappears in the MSSM vacuum right at the end of inflation~\cite{Allahverdi:2007wh}.

%%%%%%%%%%%%%%%%%%%%%%%%%%%%%%%%%%%%%%%%%%%%%%%%%%%%%%%%%%%%%%%%%%%%%%%%%%%%%%%%%%%%%%%%%%%%%%%%%%%%%%%%%%%%%%%%%%%%%%%%%%%%%%%%%%%%%%%%%%%%%%%%%%%%%%%%%%%%%%%%%%%%%%%%%%

\section{CMB Observables and Inflection point}
\label{sec-3}

In this section our primary focus is to study the cosmological observables to match the CMB data for an  inflection-point inflation
whose potential is given by Eq.~(\ref{h1a}). First we use the following parameterizations for the amplitude of the scalar 
perturbations, $P_S$, tensor perturbations, $P_T$, and the tensor-to-scalar ratio $r$, in terms of the slow-roll 
parameters~\cite{Liddle-Lyth,Liddle:1994dx}:
\begin{widetext}
\begin{eqnarray}\label{para 1}
     P_{S}(k) &=&P_{S}(k_{\star})\left(\frac{k}{k_{\star}}\right)^{n_{S}-1
+\frac{\alpha_{S}}{2}\ln\left(\frac{k}{k_{\star}}\right)+\frac{\kappa_{S}}{6}\ln^{2}\left(\frac{k}{k_{\star}}\right)+......},~~~~~\\
   \label{para 2}  P_{T}(k)&=&P_{T}(k_{\star})\left(\frac{k}{k_{\star}}\right)^{n_{T}+\frac{\alpha_{T}}{2}
\ln\left(\frac{k}{k_{\star}}\right)+\frac{\kappa_{T}}{6}\ln^{2}\left(\frac{k}{k_{\star}}\right)+......},~~~~~\\
 \label{para 3} r(k)&=&\frac{P_{T}(k)}{P_{S}(k)}=r(k_{\star})\left(\frac{k}{k_{\star}}\right)^{n_{T}-n_{S}+1+\frac{\left(\alpha_{T}-\alpha_{S}\right)}{2}\ln\left(\frac{k}{k_{\star}}\right)
+\frac{\left(\kappa_{T}-\kappa_{S}\right)}{6}\ln^{2}\left(\frac{k}{k_{\star}}\right)+....}
    \end{eqnarray}
 \end{widetext}  
where the observables are now given in terms of the inflationary potential, running of the spectral tilt $\alpha_S,~\alpha_T$, and running of the running, 
given by $\kappa_S,~\kappa_T$, where the subscripts $S$ and $T$ denote scalar and tensor modes. Here the consistency relations are modified at 
the second order due to the presence of running. The above parametrization are realized by expanding the scale dependent slow-roll parameters around the pivot scale $k=k_{\star}$ and they have been listed in an appendix ( see Eqs.~(\ref{sl1s},\ref{para 2}) ).

 %%%%%%%%%%%%%%%%%%%%%%FIGURES%%%%%%%%%%%%%%%%%%%%%%%%%%%%%%%%%%%%%
\begin{figure}[t]
{\centerline{\includegraphics[width=8.0cm, height=7.6cm] {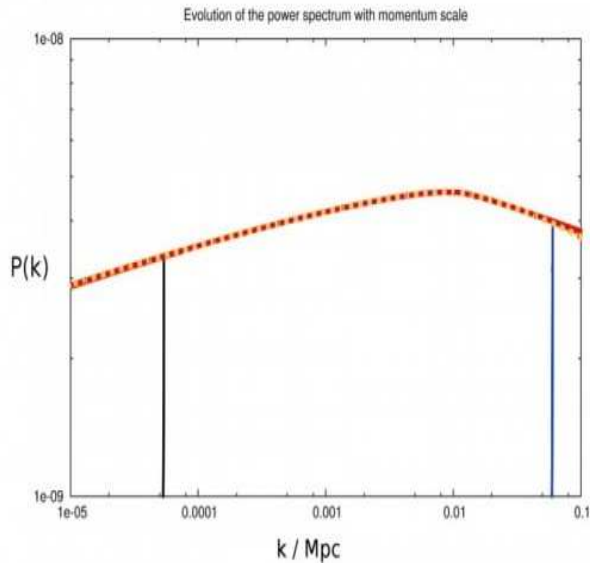}}}
\caption{We show the evolution of the power spectrum, $P(k)$, for high scale model of inflation when $H\gg m_\phi$. The 
specific values of the model parameters are:  $\delta\sim 10^{-4} $
(see Eq.~(\ref{newbeta}), $\lambda =1,~c_H=2,~a_H=2.108,~\phi_0=1.129\times 10^{16}~{\rm GeV}$
 for the entire range of momentum, $k$, that crosses the Hubble patch during the $17$ e-folds of inflation. Here the {\it orange} dotted curve is obtained from the higher order radiative corrections to the slow-roll parameters, see appendix~Eq.~(\ref{para 21a}), 
and the {\it red} curve from numerically integrating the cosmological perturbations. The ${\it blue}$ vertical line corresponds to 
$k_{max}=0.056~Mpc^{-1}$ for  $l_{max}=2500$, which is the highest probing limit of recent {\it Planck} data. On the other hand the {\it black} vertical line corresponds to $k_{min}=4.488\times 10^{-5}~Mpc^{-1}$ for $l_{min}=2$.} \label{fig2aa}
\end{figure}

%%%%%%%%%%%%%%%%%%%%%%%%%%%%%%%%%%%%%%%%%%%%%%%%%%%%%%%%%%%%%%%%%%%%%%%%%%%%%%%%%%

Cosmological parameter estimation can be done more precisely once we allow the higher order radiative corrections to the slow-roll parameters
 \cite{Peiris:2006}. We have listed  all the relevant cosmological observables in an appendix (see Eqs.~(\ref{para 21a})-(\ref{para 21i})). In our case we have obtained the predicted power spectrum from the higher order radiative corrections to the slow-roll parameters, and also numerically we have evolved the perturbations which we will discuss below.

In order to illustrate this, let us consider the case when $H\gg m_\phi$. In this case there is a possibility of detecting large tensor-scalar ratio, $r$.
We start our analysis numerically by using cosmological code CAMB~\cite{camb} with the well known 
Bunch-Davies in-in vacuum state~\cite{BD}, and also perform the quantum fluctuations via cosmological linear
perturbation theory, see Ref.~\cite{Liddle-Lyth,bookLyth}. In figure~\ref{fig2aa}, we have shown 
the evolution of the power spectrum for the entire range of momentum scale across the $17$ e-folds of inflation which has been 
observed by the Planck satellite~\cite{Planck-infl,Planck-CMBpow}. In this plot the {\it orange} dotted curve is obtained 
from the radiative corrections to the 
higher order slow-roll corrections. On the other hand the {\it red} curve is the outcome of our numerical analysis. As we can see 
that the higher order radiative corrections to the slow-roll parameters is quite a good approximation when compared to the 
results obtained from the numerical analysis for the observed range of $k_{max}={l_{max}}/{\eta_{0}\pi}\sim {\cal O}(0.056~Mpc^{-1})$ for $l_{max}=2500$ down to  $k_{min}\sim {\cal O}(10^{-5}~Mpc^{-1})$ for $l_{min}=2$, where $\eta_{0}\sim 14000~Mpc$ is the conformal time at present 
epoch~\cite{Planck-infl,Planck-CMBpow}. The slow roll approximations differ very minutely from the numerical estimation, but for very large values of 
$k$, or very small wavelength regimes beyond $l \gg 2500$.

%%%%%%%%%%%%%%%%%%%%%%%%%%%%%%%%%%%%%%%%%%%%%%%%%%%%%%%%%%%%%%%%%%%%
    
\section{Estimation of reheat temperature and the scale of inflation}\label{TRH-sec}

We need to compute the pivot scale, $k_\ast$, when the relevant perturbations had left the Hubble patch during 
inflation. We can compute by expressing the number of e-foldings during inflation,
which is given by~\cite{Planck-infl,Liddle:2003as,Burgess:2005sb}:
\begin{eqnarray}\label{efold}
N_{\star} &\approx & 71.21 - \ln \left(\frac{k_{\star}}{k_{0}}\right)  
+  \frac{1}{4}\ln{\left( \frac{V_{\star}}{M^4_{P}}\right) }\nonumber \\
&+&  \frac{1}{4}\ln{\left( \frac{V_{\star}}{\rho_{end}}\right) }  
+ \frac{1-3w_{int}}{12(1+w_{int})} 
\ln{\left(\frac{\rho_{rh}}{\rho_{end}} \right)},
\end{eqnarray}
where $\rho_{end}$ is the energy density at the end of inflation, 
$\rho_{rh}$ is an energy scale during reheating, 
$k_{0}=a_0 H_0$ is the present Hubble scale, 
$V_{\star}$ corresponds to the potential energy when the relevant modes left the Hubble patch 
during inflation and $w_{int}$ characterizes the effective equation of state 
parameter between the end of inflation and the energy scale during reheating. For our model we have $w_{int}=1/3$ exactly for
which the contribution from the last term in Eq.~(\ref{efold}) vanishes. For inflation driven by the MSSM flat directions
the time scale for the transfer of energy to the MSSM relativistic species can be computed exactly as in~\cite{Allahverdi:2011aj}. 
This happens roughly within one Hubble time. This is due to the gauge couplings of the inflaton to gauge/gaugino fields. Within $10-20$ inflaton 
oscillations radiation-dominated universe prevails, as shown in Ref.~\cite{Allahverdi:2011aj}. The resultant  {\it upper-bound} on the
reheat temperature at which all the MSSM {\it degrees of freedom} are in thermal equilibrium (kinetic and chemical equilibrium) is given by~\cite{Allahverdi:2011aj}
\begin{equation}\label{reh}
     T_{rh}=\left(\frac{30}{\pi^{2}g_{\star}}\right)^{\frac{1}{4}}\sqrt[4]{V_{\star}}\leq 6.654\times 10^{15}\sqrt[4]{\frac{r_{\star}}{0.12}}~{\rm GeV}.
\end{equation}
where we have used $g_{\star}=228.75$ (all the degrees of freedom in
MSSM). Since the temperature of the universe is so high, the lightest supersymmetric particle (LSP) relic density is then given by the 
standard (thermal) {\it freeze-out} mechanism~\cite{Jungman:1995df}. In particular, if
the neutralino is the LSP, then its relic density is determined
by its annihilation and coannihilation rates~\cite{Allahverdi:2007vy,Allahverdi:2010zp,Boehm:2012rh}.
The advantage of realizing inflation in the visible sector is that it is possible to nail down the thermal 
history of the universe precisely. At temperatures below $100$ GeV there will be no extra degrees of 
freedom in the thermal bath except that of the SM, therefore
BBN can proceed without any trouble.

For low scale models of inflation, i.e. $m_\phi \gg H$, the tensor modes are utterly negligible. For $m_\phi\sim 1$TeV,
and $\phi_0\sim 3\times 10^{14}$~GeV, the value of $H_{\ast} \sim 10^{-1}$~GeV, see Eq.(\ref{vev}). 
The estimation of the reheat temperature is given by the equality of the above 
Eq.~(\ref{reh}). The reheat temperature is typically $3\times 10^{8}$~GeV for the above parameters. Note that for $m_\phi\gg H$, 
the tensor to scalar ratio $r$ does not scale with the reheat temperature.

Note that saturating the upper-bound on $r\sim 0.12$ would yield a large reheating temperature of the universe. It is sufficiently large 
to create gravitino from a thermal bath~\cite{Ellis,Bolz:2000fu}.
The gravitino production from the direct decay of the inflaton will be suppressed~\cite{Maroto}.
In this case, the gravitino abundance is compatible with the BBN bounds, provided  the gravitino mass,
$m_{3/2}\geq {\cal O}(10)$~TeV, see~\cite{Moroi:1995fs} .The bound holds only for a decaying gravitino,  
for which the graviton will decay before the BBN. 
If gravitino happens to be the LSP,  then such a high scale model of 
inflation with large Hubble-induced corrections will be ruled out, unless there is some late entropy injection or 
there are some kinematical reasons for which the gravitino production is highly suppressed~\cite{Allahverdi:2005mz}.

%%%%%%%%%%%%%%%%%%%%%%%%%%%%%%%%%%%%%%%%%%%%%%%%%%%%%%%%%%%%%%%%%%

\section{Tensor to Scalar Ratio for high \& low scale models of inflation} \label{tensor}

The Planck constraint  implies that the tensor-to-scalar ratio, $r$, at the pivot scale $k=k_{\star}$, corresponds to an upper bound
on the energy scale of the Hubble induced inflection point inflation:
\begin{equation}\label{scale}
     V_{\star}\leq \frac{3}{2}\pi^{2}P_{S}(k_{\star})r_{\star} M^{4}_{PL}=(1.96\times 10^{16}{\rm GeV})^{4}\frac{r_{\star}}{0.12}.
   \end{equation}
For an example, at the pivot scale $k_{\star}=0.002~Mpc^{-1}$, the corresponding upper bound on the energy density 
becomes $V_\star=1.89\times 10^{16}$ GeV. The equivalent statement can be made in terms of
 the upper bound on the numerical value of the Hubble parameter during inflation as
\begin{equation}\label{hubble}
     H_{\star}\leq 9.241\times 10^{13}\sqrt{\frac{r_{\star}}{0.12}}~{\rm GeV}\,.
   \end{equation}
Here in eqn~(\ref{reh}), eqn~(\ref{scale}) and eqn~(\ref{hubble}) the equalities hold good in high scale inflationary regime ($H\gg m_{\phi}$). 
The inequalities are more significant once we enter low scale inflationary ($m_{\phi}\gg H$) region.
   %
%%%%%%%%%%%%%%%%%%%%%%%FIGURES%%%%%%%%%%%%%%%%%%%%%%%%%%%%%%

\begin{figure}[t]
{\centerline{\includegraphics[width=8cm, height=6cm] {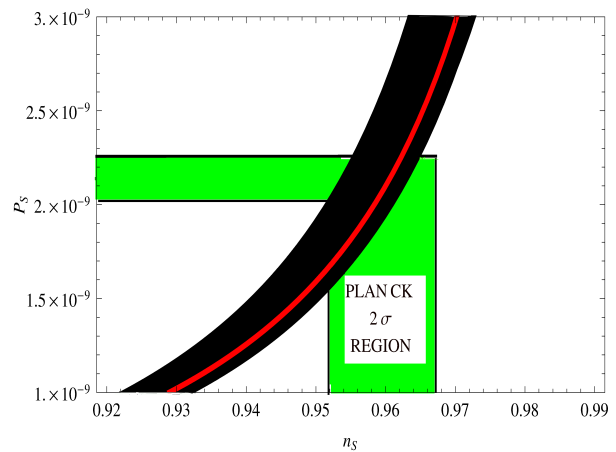}}}
\caption{ For large scale inflation, $H\gg m_\phi$, we have shown the variation of $P_{S}$~vs~$n_{S}$. The {\it red} curve shows the
model parameters, $\delta\sim 10^{-4},~\lambda =1,~c_H=2,~a_H=2.108,~\phi_0=1.129\times 10^{16}~{\rm GeV}$, for the pivot scale 
$k_{\star}=0.002~Mpc^{-1}$. The ${\it green}$ shaded region shows the  $2\sigma$ CL. range 
in $n_s$ allowed by the Planck data~\cite{Planck-infl}. Instead of getting a single solid {\it red} curve we get a {\it black} shaded region if 
we consider the full parameter space for high scale ($H\gg m_{\phi}$) inflation given by Eq.~(\ref{P-space}). } \label{fig2}
\end{figure}

%%%%%%%%%%%%%%%%%%%%%%%%%%%%%%%%%%%%%%%%%%%%%%%%%%%%%%%%%%%%
%%%%%%%%%%%%%%%%%%%%%%%%%%FIGURES%%%%%%%%%%%%%%%%%%%%%%%%%%%%%%%%%

\begin{figure}[t]
\centering
\subfigure[~r vs $n_{S}$. By varying $H_{\star}$ we can probe a wide range of tensor-to-scalar ratio: $10^{-29}< r_{\star} \leq 0.12$.
The vertical line on the left corresponds to $N=50$, while the right line corresponds to $N=70$.]{
    \includegraphics[width=8cm,height=5.3cm] {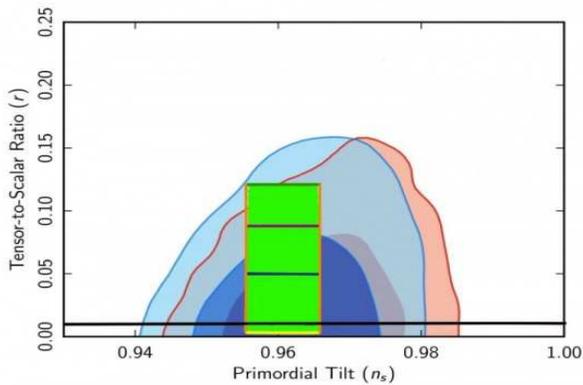}
    \label{x2}
}
\subfigure[~r vs $n_{S}$. By varying $H_{\star}$ we can probe a wide range of tensor-to-scalar ratio: $10^{-29}< r_{\star} \leq 0.12$. The vertical line on the left corresponds to $N=50$, while the right line corresponds to $N=70$.]{
    \includegraphics[width=8cm,height=5.3cm] {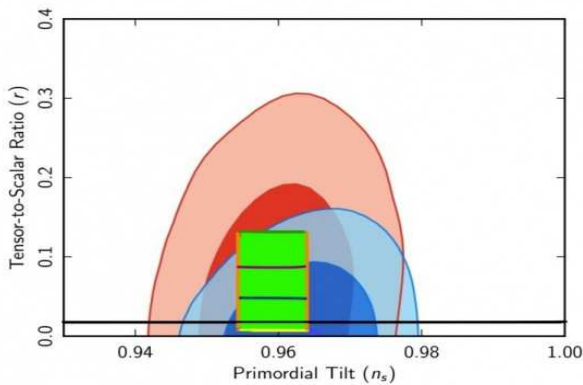}
    \label{x3}
}
\caption[Optional caption for list of figures]{ We show the joint $1\sigma$ and $2\sigma$ CL. contours using
 \subref{x2}~Planck+WMAP-9 data with~$\Lambda$CDM+r(${\it Blue~region}$), and $\Lambda$CDM+$r+\alpha_{S}$(${\it Red~region}$),
 \subref{x3}~Planck+WMAP-9+BAO data with~$\Lambda$CDM+r(${\it Blue~region}$) and $\Lambda$CDM+$r+\alpha_{S}$(${\it Red~region}$).
The straight lines parallel to $n_{S}$ axis are drawn by varying the Hubble parameter $H_{\star}$ within
the range $10^{-1}~{\rm GeV}< H_{\star} \leq 9.241\times 10^{13}~{\rm GeV}$.
 The {\it deep green} line and the {\it yellow } line correspond to the upper and lower bound of $H_{\star}$ respectively. The {\it green}
 shaded region bounded by {\it orange} lines represent the allowed region obtained from the model.
 Additionally, the {\it black} thick line divides the low scale ($m_{\phi}\gg H$) and the
high scale ($H\gg m_{\phi}$) regions of inflation. 
}
\label{fig3}
\end{figure}

%%%%%%%%%%%%%%%%%%%%%%%%%%FIGURES%%%%%%%%%%%%%%%%%%%%%%%%%%%%%%%%%

\begin{figure}[t]
\centering
\subfigure[~$\alpha_{S}$ vs $n_{S}$]{
    \includegraphics[width=8cm,height=5.3cm] {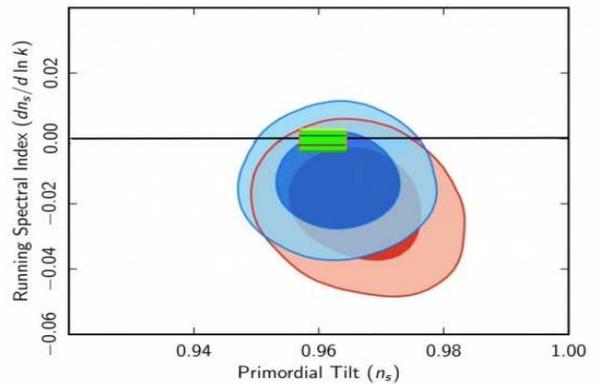}
    \label{z1}}
\subfigure[~$\kappa_{S}$ vs $\alpha_{S}$]{
    \includegraphics[width=8cm,height=5.3cm] {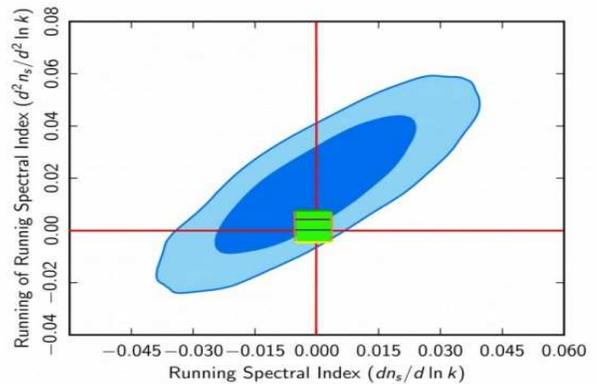}
    \label{z2}
}
\caption[Optional caption for list of figures]{ We show the joint $1\sigma$ and $2\sigma$ CL. contours 
using {\it Planck}+WMAP-9+BAO with \subref{z1}~$\Lambda$CDM+$\alpha_{S}$(${\it Blue~ region}$) and 
$\Lambda$CDM+$\alpha_{S}+r$(${\it Red~ region}$),
\subref{z2}~$\Lambda$CDM+$\alpha_{S}+\kappa_{S}$(${\it Blue~ region}$) background. 
The straight lines parallel to $n_{S}$ axis are drawn by varying the Hubble parameter $H_{\star}$ within
the range $10^{-1}~{\rm GeV}< H_{\star} \leq 9.241\times 10^{13}~{\rm GeV}$.
 The {\it deep green} line and the {\it yellow } line correspond to the upper and lower bound of $H_{\star}$ respectively. The {\it green}
 shaded region bounded by {\it orange} lines represent the allowed region obtained from the model.
}
\label{fig4}
\end{figure}

%%%%%%%%%%%%%%%%%%%%%%%%%%%%%%%%%%%%%%%%%%%%%%%%%%%%%%%%%%%%%%%%%
We scan the model parameters for obtaining large tensor to scalar ratio, $r$, for the following values:
\begin{eqnarray}\label{P-space}
c_{H} &\sim &{\cal O}(1-10)\,, \nonumber \\
a_{H} & \sim &{\cal O}(10-100)\,, \nonumber \\
\lambda &\sim & {\cal O}(1)\,, \nonumber \\
\phi_{0} &\sim & {\cal O}((1-3)\times 10^{16}GeV)
\end{eqnarray}
Now including the higher order corrections to the slow-roll parameters, 
the inflationary observables are estimated from our model as following:
\begin{eqnarray}
2.092<10^{9}P_{S}<2.297\,, \nonumber \\
0.958<n_{S}<0.963\,, \nonumber \\
%0.010<n_{T}<0.015\,, \nonumber \\
r<0.12\,, \nonumber \\
-0.0098<\alpha_{S}<0.0003\,, \nonumber \\
%0.00002<\alpha_{T}<0.0004\,, \nonumber \\
-0.0007<\kappa_{S}<0.006\,
%0.000007<\kappa_{T}<0.00003 
\end{eqnarray}

which confronts the {\it Planck}+WMAP-9+BAO data set, \cite{Planck-infl,Planck-1,Planck-CMBpow,Planck-datans}, well within $2\sigma$ CL.
Furthermore, we consider the following values of the model parameters which match the TT-spectrum of the CMB data for high scale model of 
inflation, i.e. $H\gg m_\phi$,
 \begin{eqnarray}\label{paraspace}
&&\delta\sim 10^{-4},~~~\lambda =1,~~~c_H=2,~~~a_H=12.650,\nonumber \\
&& \phi_0=1.129\times 10^{16}~{\rm GeV}=4.704\times 10^{-3}M_{PL}.
 \end{eqnarray}

Using Eq.~(\ref{paraspace}), in Fig.~(\ref{fig2}) we have shown the behavior of the amplitude of the the power spectrum, $P_{S}$ with respect to spectral tilt, $n_{S}$
 at the pivot scale $k_{\star}=0.002~{\rm Mpc}^{-1}$ by a {\it red} curve. If we take care of the full parameter
space, see Eq.~(\ref{P-space}), there are solutions which have been shown in a black shaded region.

 In principle, we can vary $H_\ast$ from high scales to low scales. Since in our case the advantage is that the thermal history is well established, 
 we can trace the relevant number of e-foldings, i.e . $N_\ast$ for various ranges of $H_\ast$.
 By varying $10^{-1}~{\rm GeV}< H_{\star} \leq 9.241\times 10^{13}~{\rm GeV}$, we can probe tensor-to-scalar ratio for a wide range: $10^{-29}< r_{\star} \leq 0.12$.

 Furthermore, by using {\it Planck}+WMAP-9 and {\it Planck}+WMAP-9+BAO datasets
 with $\Lambda$CDM background
 along with different  combined constraints, we have shown the status of inflection point inflationary model in the 
 marginalized $1\sigma$ and $2\sigma$ CL.
 contours  in Fig.~(\ref{fig3}). The allowed region from the model is explicitly shown by the {\it green} shaded
 region bounded by {\it orange} vertical lines parallel to $r$-axis. Along the vertical lines 
the number of e-foldings varies within $50<N_{\star}<70$ (from left to right) for various ranges of $H_\ast$ as mentioned earlier.
 We have also shown a thick {\it black} line parallel to $n_{S}$ axis in Fig.~(\ref{fig3}) which 
discriminates between the low scale $(m_{\phi}\geq  H)$ and high scale $(H\gg m_{\phi})$ inflationary scenarios. 
Additionally, we have depicted various straight 
lines for the intermediate values of $H_{\star}$ within the allowed region. The model also provides very mild running, $\alpha_{s}$, 
 and running of running, $\kappa_{S}$, which is also shown in the marginalized $1\sigma$ and $2\sigma$ CL. contours in Fig.~(\ref{fig4}), 
 where we have  used {\it Planck}+WMAP-9 and {\it Planck}+WMAP-9+BAO datasets with $\Lambda$CDM background along with different combined observational constraints.

%%%%%%%%%%%%%%%%%%%%%%%%%%%%%%%%%%%%%%%%%%%%%%%%%%%%%%%%%%%%%%%%%%%%%%%%%%%%%%%%%%%%%%%%%%%%%%%%%%%%%%%%%%%%%%%%%%%%%%%%%%%%%%%%%%%%%%%%%%%%%%%%%%%%%%%%%%%%%%%%%%%%%%%%%%%%%%%%%%%%%%%%%%%%%

\section{\bf Multipole scanning of TT-spectra and CMB anisotropy}\label{sec-4}

In this section we study the TT-angular power spectrum for the CMB anisotropy. For our present setup 
at low $\ell$ region $(2<l<49)$ the contributions from the running ($\alpha_{S},\alpha_{T}$), and running of running ($\kappa_{S},\kappa_{T}$) are very small.
Consequently their additional contribution to the power spectrum for scalar and tensor modes becomes unity $(\sim {\cal O}(1))$ (this is consistent with 
the initial condition at the pivot scale $k=k_{\star}$) and the original
 power spectrum becomes unchanged. As a result the proposed model will be well fitted with the {\it Planck } low-$l$ data within 
 high cosmic variance except for a few outliers. On the other hand, when we move
 towards high $\ell$ regime ($50<l<2500$) the contribution of running and running of running become stronger and
 this will enhance the power spectrum to a permissible value such that it will accurately fit {\it Planck} high-$l$ 
data within very small cosmic variance. In this way one can easily survey over all the multipoles starting from low-$l$ to high-$l$ using the
same parameterizations as mentioned in Eqs.~(\ref{paraspace}).

%%%%%%%%%%%%%%%%%%FIGURE%%%%%%%%%%%%%%%%%%%%%%%%%%%%%%%%%

\begin{figure}[t]
\centering
\includegraphics[width=8.8cm,height=7.5cm]{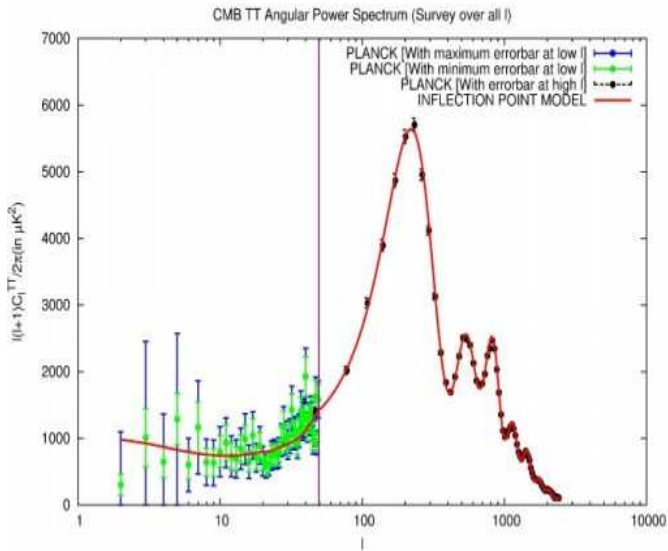}
\caption{TT-power spectrum for $\ell$~($2<l<2500$). The vertical line is drawn at $l=50$ which separates the
 low-$l$ ($2<l<50$) and high-$l$ ($50<l<2500$) region. Here the TT-power spectrum is drawn for the parameter values mentioned
in Eq.~(\ref{paraspace}) in the context of high scale ($H\gg m_{\phi}$) Hubble induced inflationary framework.}
\label{fig6}
\end{figure}

%%%%%%%%%%%%%%%%%%%%%%%%%%%%%%%%%%%%%%%%%%%%%%%%%%%%%%%%%%%%%%

From Fgs.~(\ref{fig6}), we see that the Sachs-Wolfe plateau obtained
from our model is non flat, confirming the appearance of running, and running of the running in the spectrum 
observed for low $l$ region ($l<50$).
 For larger value of the multipole ($50<l<2500$),
CMB anisotropy spectrum is dominated by the Baryon
Acoustic Oscillations (BAO), giving rise to several ups and
downs in the spectrum, see~\cite{Hu:1996qs}.   Note  that high $l$ regions of our model are well fitted within the 
small cosmic variance observed by {\it Planck}. In the low $l$ region due to the presence of very 
large cosmic variance there may be other pre-inflationary scenarios which might be able to fit the TT-power spectrum better~\cite{Planck-CMBpow}. 
In our study we have considered only the possibility for which the model is well fitted for both 
low and high $l$ regions.

%%%%%%%%%%%%%%%%%%%%%%%%%%%%%%%%%%%%%%%%%%%%%%%%%%%%%%%%%%%%%%%%%%%%%%%%%%%%%%%%%%%%%%%%%%%%%%%%%%%%%%%%%%%%%%%%%%%%%%%%%%%%%%%%%%%%%%%%%%%%%%%%%%%%%%%%%%%%
 
\section{Conclusion}\label{sec-5}

In this paper we have considered a simple model of inflection-point inflation motivated by the MSSM flat directions, where 
we have taken potentials with both supergravity corrections are important and negligible. In the former case,
we yield significantly large tensor-to-scalar ratio, $r\leq 0.12$, for $H_\star\sim 9.24\times 10^{13}$~GeV and 
the VEV $\phi_0 =1.12\times 10^{16}$~GeV. 
The model fits the amplitude of the power spectrum and the spectral tilt. The model predicts mild running and running of the running 
of the spectral tilt well within the $2\sigma$ uncertainties. In particularly, the high scale inflection-point inflation model fits the high 
$l$ multipoles of the {\it Planck} data quite well with the $\Lambda$CDM parameters. The low $l$ multipoles have high uncertainties and they are
within the cosmic variance. The forthcoming polarization data from
Planck will hopefully further constrain the inflection-point model of inflation.  The model tends to predict a perfect match for the spectral tilt
even for a small tensor-to-scalar ratio, $r$, as seen in Fig.~(\ref{fig3}). 

The perturbations created from the slow roll evolution of the inflaton are Gaussian and adiabatic. The amplitude and the spectral tilt  match 
very well with the {\it Planck} data. One of the advantages of the proposed model is that it is embedded fully within MSSM, and therefore it predicts 
the right thermal history of the universe with {\it no extra relativistic degrees of freedom} other than that of the  Standard Model.

%%%%%%%%%%%%%%%%%%%%%%%%%%%%%%%%%%%%%%%%%%%%%%%%%%%%%%%%%%%%%%%%%%%%%%%%%%%%%%%%%%%%%%%%%%%%%%%%%%%%%%%%%%%%%%%%%%%%%%%%%%%%%%%%%%%%%%%%%%%%%%%%%%%%%%%%%%%%%%%%%%%%%%%%%%%%%%%%%%%%%%%%%%%%%%%%%%%%%%%%

\section*{Acknowledgments}

AM would like to thank Pratika Dayal and Seshadri Nadathur for helpful discussions. 
SC thanks Council of Scientific and
Industrial Research, India for financial support through Senior
Research Fellowship (Grant No. 09/093(0132)/2010). SC also thanks Arnab Dasgupta for useful suggestions. AM is supported 
by the Lancaster-Manchester-Sheffield Consortium for Fundamental Physics under STFC grant ST/J000418/1.
SP thanks ISI Kolkata for computational support through a research grant.

%%%%%%%%%%%%%%%%%%%%%%%%%%%%%%%%%%%%%%%%%%%%%%%%%%%%%%%%%%%%%%%%%%%%%%%%%%%%%%%%%%%%%%%%%%%%%%%%%%%%%%%%%%%%%%%%%%%%%%%%%%%%%%%%%%%%%%%%%%%%%%%%%%%%%%%%%%%%%%%%%%%%%%%%%%%%%%%%%%%%%%%%%%%%%%%%%%%%%%%%%%%%%%%%%%%%%%%%%
\section*{\bf Appendix}

The power spectrums as well as other relevant parameters in terms of the potential can be written as~\cite{Liddle-Lyth,Liddle:1994dx}:
\begin{widetext}
\be\begin{array}{llll}\label{sl1s}
    \displaystyle \epsilon_{V}(k)=\epsilon_{V}-\frac{\alpha_{T}}{2}\ln\left(\frac{k}{k_{\star}}\right)+\frac{\kappa_{T}}{4}\ln^{2}\left(\frac{k}{k_{\star}}\right)+.......,\\
 \displaystyle \eta_{V}(k)=\eta_{V}-\frac{\left(\alpha_{S}-3\alpha_{T}\right)}{2}\ln\left(\frac{k}{k_{\star}}\right)+\frac{1}{2}\left(
\kappa_{S}-3\kappa_{T}+\left\{n^{2}_{T}+\alpha_{T}\right\}\left[\frac{n_{S}-3n_{T}-1}{2}\right]\right.\\ \left.~~~~~~~~~~~~~~~~~~~~~~~~~~~~~~~~~~~
~~~~~~~~~~~~~~~~~~~~~~~~~~~~~~~\displaystyle +\frac{\xi^{2}_{V}}{2}\left\{1-n_{S}-3n_{T}\right\}
\right)\ln^{2}\left(\frac{k}{k_{\star}}\right)+.......,\\
 \displaystyle \xi^{2}_{V}(k)=\xi^{2}_{V}-\frac{1}{2}\left(\kappa_{S}-4\kappa_{T}+4n^{2}_{T}\left\{n_{S}-n_{T}-1
\right\}\right)\ln\left(\frac{k}{k_{\star}}\right)\\ ~~~~~~~~~~~~~~~~~~~~~~~~~~~~~~~~~~
\displaystyle +\frac{1}{4}\left(\xi^{2}_{V}\left\{16n^{2}_{T}+9n_{T}\left[n_{S}-3n_{T}-1\right]+\left[n_{S}-3n_{T}-1\right]^{2}+2\xi^{2}_{V}
\right\}\right)\ln^{2}\left(\frac{k}{k_{\star}}\right)+.......,\\
 \displaystyle \sigma^{3}_{V}(k)=\sigma^{3}_{V}+\sigma^{3}_{V}\left(1-n_{S}\right)
\ln\left(\frac{k}{k_{\star}}\right)+\frac{\sigma^{3}_{V}}{4}\left(30n^{2}_{T}+20n_{T}\left[n_{S}-3n_{T}-1\right]+\xi^{2}_{V}+2\left[n_{S}-3n_{T}-1\right]^{2}
\right)\ln^{2}\left(\frac{k}{k_{\star}}\right)+...... 
   \end{array}\ee

where at the pivot scale $k=k_{\star}$ the slow roll parameters are defined as follows:
    \begin{eqnarray}\label{para 2}
     \epsilon_{V}=\frac{M^{2}_{P}}{2}\left(\frac{V^{\prime}}{V}\right)^{2}, ~~~~~
     \eta_{V}={M^{2}_{P}}\left(\frac{V^{\prime\prime}}{V}\right),~~~~~
     \xi^{2}_{V}=M^{4}_{P}\left(\frac{V^{\prime}V^{\prime\prime\prime}}{V^{2}}\right),~~~~~
     \sigma^{3}_{V}=M^{6}_{P}\left(\frac{V^{\prime 2}V^{\prime\prime\prime\prime}}{V^{3}}\right).
    \end{eqnarray}
\end{widetext}
Additionally, the appearance of  very mild running ($\alpha_{S},\alpha_{T}$) and running of the running ($\kappa_{S},\kappa_{T}$) of the spectrum 
might have implications for the {\it Primordial black hole } formation \cite{Drees:2011yz}.

 %Eventually, we are able to express the power spectrums as well as the other relevant parameters in terms of the potential~\cite{Liddle:1994dx,Liddle-Lyth,ChoudhuryPRD,Choudhuryar1,Choudhuryar2,ChoudhuryJCS}:
%\begin{widetext}
%\begin{eqnarray}\label{para 21}
%    P_{S}(k_{\star}) &= &\frac{V}{24\pi^{2}M^{4}_{PL}\epsilon_{V}},~~~~~
%   \label{para 4}  P_{T}(k_{\star})=\frac{2V}{3\pi^{2}M^{4}_{PL}} \\
 %   \label{para 6} n_{S}-1&=&\frac{d\ln P_{S}}{d\ln k}=2\eta_{V}-6\epsilon_{V},~~~~~
%   \label{para 8}  n_{T}=\frac{d\ln P_{T}}{d\ln k}=-2\epsilon_{V},\\
 %  \label{para 10}  r(k_{\star})&=&\frac{P_{T}(k_{\star})}{P_{S}(k_{\star})}=16\epsilon_{V},\\
%   \label{para 12}  \alpha_{S}&=&\frac{dn_{S}}{d\ln k}=16\eta_{V}\epsilon_{V}-24\epsilon^{2}_{V}-2\xi^{2}_{V},~~~~~
 %\label{para 14}\alpha_{T}=\frac{dn_{T}}{d\ln k}=4\eta_{V}\epsilon_{V}-8\epsilon^{2}_{V},\\
%    \label{para 16} \kappa_{S}&=&\frac{d^{2}n_{S}}{d\ln k}=192\epsilon^{2}_{V}\eta_{V}-192\epsilon^{3}_{V}+2\sigma^{3}_{V}
%-24\epsilon_{V}\xi^{2}_{V}+2\eta_{V}\xi^{2}_{V}-32\eta^{2}_{V}\epsilon_{V}\\
%    \label{para 181} \kappa_{T}&=&\frac{d^{2}n_{T}}{d\ln k}=56\eta_{V}\epsilon^{2}_{V}-64\epsilon^{3}_{V}
%-8\eta^{2}_{V}\epsilon_{V}-4\epsilon_{V}\xi^{2}_{V}.
 %   \end{eqnarray}
%\end{widetext}
Once the higher order radiative corrections are incorporated due to the presence of running in the parameterization of the power spectrum, then 
the inflationary observables can be expressed as:
\label{appendix}
\begin{widetext}
  \be\begin{array}{lllll}\label{para 21a} \displaystyle P_{S}(k_{\star}) 
=\left[1-(2{\cal C}_{E}+1)\epsilon_{V}+{\cal C}_{E}\eta_{V}\right]^{2}\frac{V}{24\pi^{2}M^{4}_{PL}\epsilon_{V}},%\end{array}\ee
  %\be\begin{array}{lllll}\label{para 21b} 
~~~~\displaystyle  P_{T}(k_{\star})=\left[1-({\cal C}_{E}+1)\epsilon_{V}\right]^{2}\frac{2V}{3\pi^{2}M^{4}_{PL}}, \end{array}\ee
  \be\begin{array}{lllll}\label{para 21c} \displaystyle  n_{S}-1%&=&\frac{1}{(1-\frac{\epsilon_{V}}{3}-\frac{\eta_{V}}{3})}\left\{(2\eta_{V}
%-6\epsilon_{V})-\frac{\left[2{\cal C}_{E}\xi^{2}_{V}-8(2{\cal C}_{E}+1)\epsilon^{2}_{V}
%-4\epsilon_{V}\eta_{V}(3{\cal C}_{E}+1)\right]}{\left[1-(2{\cal C}_{E}+1)\epsilon_{V}+{\cal C}_{E}\eta_{V}\right]}\right\},\\
     %\nonumber
 \approx (2\eta_{V}
-6\epsilon_{V})-2{\cal C}_{E}\xi^{2}_{V}+\frac{2}{3}\eta^{2}_{V}+2(8{\cal C}_{E}+3)\epsilon^{2}_{V}
+2\epsilon_{V}\eta_{V}\left(6{\cal C}_{E}+\frac{7}{3}\right)%\\ \displaystyle~~~~~~~~~~~~~~~~~~~~~~~~~~~~~~~~~~~~~~~~~~~~~~~~~~~~~~~~~~~~~~~~~~~~~~
%~~~~~~~
-4{\cal C}_{E}({\cal C}_{E}+1)\xi^{2}_{V}\epsilon_{V}+2C^{2}_{E}\eta_{V}\xi^{2}_{V},\end{array}\ee
  \be\begin{array}{lllll}\label{para 21d}  \displaystyle n_{T}%&=&\frac{1}{(1-\frac{\epsilon_{V}}{3}-\frac{\eta_{V}}{3})}\left\{-2\epsilon_{V}+\frac{4({\cal C}_{E}+1)(\epsilon_{V}\eta_{V}-2\epsilon^{2}_{V})}
%{\left[1-(2{\cal C}_{E}+1)\epsilon_{V}\right]}\right\}\\ \nonumber
\approx-2\epsilon_{V}+2\left(2{\cal C}_{E}+\frac{5}{3}\right)\epsilon_{V}\eta_{V}-2\left(4{\cal C}_{E}+\frac{13}{3}\right)\epsilon^{2}_{V},\end{array}\ee
 \be\begin{array}{lllll}\label{para 21e} \displaystyle  r(k_{\star})=16\epsilon_{V}\frac{\left[1-({\cal C}_{E}+1)\epsilon_{V}\right]^{2}}{\left[1-(2{\cal C}_{E}+1)\epsilon_{V}
+{\cal C}_{E}\eta_{V}\right]^{2}}\approx16\epsilon_{V}\left[1+2{\cal C}_{E}(\epsilon_{V}-\eta_{V})\right],\end{array}\ee
  \be\begin{array}{lllll}\label{para 21f}  \displaystyle \alpha_{S}\approx\left(16\eta_{V}\epsilon_{V}-24\epsilon^{2}_{V}-2\xi^{2}_{V}\right)-2{\cal C}_{E}(4\epsilon_{V}\xi^{2}_{V}
-\eta_{V}\xi^{2}_{V}-\sigma^{3}_{V})+\frac{4}{3}\eta_{V}(2\eta_{V}\epsilon_{V}-\xi^{2}_{V})
 %\\ ~~~~~~~~~
\displaystyle+4(8{\cal C}_{E}+3)\epsilon_{V}(4\epsilon^{2}_{V}-2\eta_{V}\epsilon_{V})
 
\\ \displaystyle~~~~~~~~~
-4{\cal C}_{E}({\cal C}_{E}+1)\left[\epsilon_{V}(4\epsilon_{V}\xi^{2}_{V}-\eta_{V}\xi^{2}_{V}-\sigma^{3}_{V})+\xi^{2}_{V}(4\epsilon^{2}_{V}-2\eta_{V}\epsilon_{V})\right]
+2{\cal C}^{2}_{E}\xi^{2}_{V}(2\eta_{V}\epsilon_{V}-\xi^{2}_{V})\\  \displaystyle
~~~~~~~~~+2{\cal C}^{2}_{E}\eta_{V}(4\epsilon_{V}\xi^{2}_{V}-\eta_{V}\xi^{2}_{V}-\sigma^{3}_{V}),\end{array}\ee
\be\begin{array}{lllll}\label{para 21g} \displaystyle\alpha_{T}\approx(4\eta_{V}\epsilon_{V}-8\epsilon^{2}_{V})+2\left(2{\cal C}_{E}
+\frac{5}{3}\right)\left[\epsilon_{V}(2\eta_{V}\epsilon_{V}-\xi^{2}_{V})
+\eta_{V}(4\epsilon^{2}_{V}-2\eta_{V}\epsilon_{V})\right]%\\ \displaystyle~~~~~~~~~
-4\left(4{\cal C}_{E}+\frac{13}{3}\right)
\epsilon_{V}(4\epsilon^{2}_{V}-2\eta_{V}\epsilon_{V}),\end{array}\ee
 \be\begin{array}{lllll}\label{para 21h} \displaystyle   \kappa_{S}\approx192\epsilon^{2}_{V}\eta_{V}-192\epsilon^{3}_{V}+2\sigma^{3}_{V}
-24\epsilon_{V}\xi^{2}_{V}+2\eta_{V}\xi^{2}_{V}-32\eta^{2}_{V}\epsilon_{V}-8{\cal C}_{E}\left[\epsilon_{V}(4\epsilon_{V}\xi^{2}_{V}-\eta_{V}\xi^{2}_{V}-\sigma^{3}_{V})
+\xi^{2}_{V}(4\epsilon^{2}_{V}-2\eta_{V}\epsilon_{V})\right]
\\~~~~~~~~~\displaystyle +2{\cal C}_{E}\left[\eta_{V}(4\epsilon_{V}\xi^{2}_{V}-\eta_{V}\xi^{2}_{V}-\sigma^{3}_{V})
+\xi^{2}_{V}(2\eta_{V}\epsilon_{V}-\xi^{2}_{V})\right]+4{\cal C}_{E}\sigma^{3}_{V}(3\epsilon_{V}-\eta_{V})+\frac{4}{3}(2\eta_{V}\epsilon_{V}-\xi^{2}_{V})^{2}\\
~~~~~~~~~\displaystyle +\frac{4}{3}\eta_{V}\left[2\eta_{V}(4\epsilon^{2}_{V}-2\eta_{V}\epsilon_{V})+2\epsilon_{V}(2\eta_{V}\epsilon_{V}-\xi^{2}_{V})-(4\epsilon_{V}\xi^{2}_{V}-\eta_{V}\xi^{2}_{V}-\sigma^{3}_{V})\right] 
 \\ ~~~~~~~~~\displaystyle+4(8{\cal C}_{E}+3)(4\epsilon^{2}_{V}-2\eta_{V}\epsilon_{V})^{2}+16(8{\cal C}_{E}+3)\epsilon_{V}\left[(2\epsilon_{V}-\eta_{V})(4\epsilon^{2}_{V}-2\eta_{V}\epsilon_{V})
-\epsilon_{V}(2\eta_{V}\epsilon_{V}-\xi^{2}_{V})\right]
\\ \displaystyle ~~~~~~~~~+4\left(6{\cal C}_{E}+\frac{7}{3}\right)(2\eta_{V}\epsilon_{V}-\xi^{2}_{V})(4\epsilon^{2}_{V}-2\eta_{V}\epsilon_{V})
+2\left(6{\cal C}_{E}+\frac{7}{3}\right)\epsilon_{V}\left[2(2\eta_{V}\epsilon_{V}-\xi^{2}_{V})\epsilon_{V}+2\eta_{V}(4\epsilon^{2}_{V}-2\eta_{V}\epsilon_{V})
\right.\\ \left.\displaystyle ~~~~~~~~~-(4\epsilon_{V}\xi^{2}_{V}
-\eta_{V}\xi^{2}_{V}-\sigma^{3}_{V})\right]-4{\cal C}_{E}({\cal C}_{E}+1)(4\epsilon^{2}_{V}-2\eta_{V}\epsilon_{V})(4\epsilon_{V}\xi^{2}_{V}-\eta_{V}\xi^{2}_{V}-\sigma^{3}_{V})
\\ \displaystyle~~~~~~~~~
-4{\cal C}_{E}({\cal C}_{E}+1)\epsilon_{V}\left[(4\epsilon_{V}-\eta_{V})(4\epsilon_{V}\xi^{2}_{V}-\eta_{V}\xi^{2}_{V}-\sigma^{3}_{V})+(16\epsilon^{2}_{V}+\xi^{2}_{V}
-10\eta_{V}\epsilon_{V})\xi^{2}_{V}-2\sigma^{3}_{V}(3\epsilon_{V}-\eta_{V})\right]
\\ \displaystyle~~~~~~~~~
-4{\cal C}_{E}({\cal C}_{E}+1)\left[(4\epsilon_{V}\xi^{2}_{V}-\eta_{V}\xi^{2}_{V}-\sigma^{3}_{V})(4\epsilon^{2}_{V}-2\eta_{V}\epsilon_{V})
+2\xi^{2}_{V}((4\epsilon_{V}-\eta_{V})[4\epsilon^{2}_{V}-2\eta_{V}\epsilon_{V}]-\epsilon_{V}[2\eta_{V}\epsilon_{V}-\xi^{2}_{V}])\right]
\\ \displaystyle ~~~~~~~~~+2{\cal C}^{2}_{E}[(4\epsilon_{V}\xi^{2}_{V}-\eta_{V}\xi^{2}_{V}-\sigma^{3}_{V})(2\eta_{V}\epsilon_{V}-\xi^{2}_{V})
+\xi^{2}_{V}(2\eta_{V}(4\epsilon^{2}_{V}-2\eta_{V}\epsilon_{V})+2\epsilon_{V}(2\eta_{V}\epsilon_{V}-\xi^{2}_{V})-[4\epsilon_{V}\xi^{2}_{V}-\eta_{V}\xi^{2}_{V}
\\~~~~~~~~~-\sigma^{3}_{V}])]+2{\cal C}^{2}_{E}[(2\eta_{V}\epsilon_{V}-\xi^{2}_{V})
(4\epsilon_{V}\xi^{2}_{V}-\eta_{V}\xi^{2}_{V}-\sigma^{3}_{V})+\eta_{V}([4\epsilon_{V}-\eta_{V}]
[4\epsilon_{V}\xi^{2}_{V}-\eta_{V}\xi^{2}_{V}-\sigma^{3}_{V}]+\xi^{2}_{V}[16\epsilon^{2}_{V}+\xi^{2}_{V}\\ \displaystyle ~~~~~~~~~
-10\eta_{V}\epsilon_{V}]-2\sigma^{3}_{V}[3\epsilon_{V}-\eta_{V}])],\end{array}\ee
 \be\begin{array}{lllll}\label{para 21i}  \kappa_{T}\approx 56\eta_{V}\epsilon^{2}_{V}-64\epsilon^{3}_{V}
-8\eta^{2}_{V}\epsilon_{V}-4\epsilon_{V}\xi^{2}_{V}+2\left(2{\cal C}_{E}
+\frac{5}{3}\right)\left[(2\eta_{V}\epsilon_{V}-\xi^{2}_{V})
(4\epsilon^{2}_{V}-2\eta_{V}\epsilon_{V})\right.\\ \left.~~~~~~~~~\displaystyle 
+\epsilon_{V}(2\eta_{V}[4\epsilon^{2}_{V}-2\eta_{V}\epsilon_{V}]+2\epsilon_{V}[2\eta_{V}\epsilon_{V}-\xi^{2}_{V}]-
[4\epsilon_{V}\xi^{2}_{V}-\eta_{V}\xi^{2}_{V}-\sigma^{3}_{V}])
\right.\\ \left.~~~~~~~~~+\eta_{V}(8\epsilon_{V}[4\epsilon^{2}_{V}-2\eta_{V}\epsilon_{V}]-2\eta_{V}[4\epsilon^{2}_{V}-2\eta_{V}\epsilon_{V}]
-2\epsilon_{V}[2\eta_{V}\epsilon_{V}-\xi^{2}_{V}])\right]\\ \displaystyle~~~~~~~~~~
-4\left(4{\cal C}_{E}+\frac{13}{3}\right)[(4\epsilon^{2}_{V}-2\eta_{V}\epsilon_{V})^{2}+\epsilon_{V}(
(8\epsilon_{V}-2\eta_{V})[4\epsilon^{2}_{V}-\epsilon_{V}]-2\epsilon_{V}[2\eta_{V}\epsilon_{V}-\xi^{2}_{V}])].
\end{array}\ee
\end{widetext}

where ${\cal C}_{E}=4(\ln 2+\gamma_{E})-5$ with $\gamma_{E}=0.5772$ is the {\it Euler-Mascheroni constant}
 originating in the expansion
of the gamma function. For details discussion on these aspects see Refs~\cite{Peiris:2006}.

%%%%%%%%%%%%%%%%%%%%%%%%%%%%%%%%%%%%%%%%%%%%%%%%%%%%%%%%%%%%%%%%%%%%%%%%%%%%%%%%%%%%%%%%%%%%%%%%%%%%%%%%%%%%%%%%%%%%%%%%%%%%%%%%%%%%%%%%%%%%%%%%%%%%%%%%%%%%%%%%%%%%%%%%%%%%%%%%%%%%%%%%%%%%%%%%%%%%%%%%%%%%%%%%%%%%%%%%%%%%%%%%%%%%

\end{document}